\documentclass[aps,prd,nofootinbib]{revtex4}

\usepackage{amsmath,amssymb,amsfonts,epsfig}
\usepackage{color,graphicx}
\usepackage{subfigure}
\usepackage{bm}

\newcommand{\be}{\begin{equation}}
\newcommand{\ee}{\end{equation}}
\newcommand{\bea}{\begin{eqnarray}\displaystyle}
\newcommand{\eea}{\end{eqnarray}}

\newcommand{\bhat}[1]{\hat{\mathbf{#1}}}

\def\d{{\rm d}}
\def\b{\mathbf}
\def\LCDM{$\Lambda$CDM}

\newcommand{\lb}{\left(}
\newcommand{\rb}{\right)}
\newcommand{\intk}{\int\frac{\mathrm{d}^3k}{(2\pi)^3}}
\newcommand{\stn}{\frac{\text{S}}{\text{N}}}
\newcommand{\la}{\left\langle}
\newcommand{\ra}{\right\rangle}

\begin{document}

\title{Searching for Stringy Topologies in the Cosmic Microwave Background}

\author{Assaf Ben-David}
\email{bd.assaf@gmail.com}
\author{Ben Rathaus}
\email{ben.rathaus@gmail.com}
\author{Nissan Itzhaki}
\email{nitzhaki@post.tau.ac.il}

\affiliation{Raymond and Beverly Sackler Faculty of Exact Sciences,
School of Physics and Astronomy, Tel-Aviv University, Ramat-Aviv, 69978,
Israel}

\begin{abstract}

We consider a universe with a non-classical stringy topology that has fixed points. We concentrate on the simplest example, an orbifold point, and study its observable imprints on the cosmic microwave background (CMB). We show that an orbifold preserves the Gaussian nature of the temperature fluctuations, yet modifies the angular correlation function. A direct signature of an orbifold is a single circle in the CMB that is invariant under rotation by $180^\circ$. Searching the 7-year ILC map of WMAP, we find one candidate circle with high statistical significance. However, a closer look reveals that the temperature profile does not fit an orbifold. We place a lower bound on the distance to an orbifold point at $\sim85\%$ of the distance to the surface of last scattering. 

\end{abstract}

\maketitle

\section{Introduction}

There is a  strong experimental evidence that our universe is, to a good approximation, flat \cite{Komatsu:2010fb}. This, however, does not necessarily mean that the  topology of the universe is $\mathbb{R}^3$. A non-trivial topology is a fascinating possibility that was investigated via its imprints on the cosmic microwave background (CMB) quite extensively (see e.g. \cite{Starobinsky:1993yx,deOliveiraCosta:1995td,Cornish:1997rp,Park:1997ad,Cornish:1997ab,Cornish:2003db,Phillips:2004nc,ShapiroKey:2006hm,Aurich:2010wf,Bielewicz:2010bh,Vaudrevange:2012da}). So far the focus  was on classical topologies \cite{Riazuelo:2003ud}. Such topologies can be viewed as $\mathbb{R}^3$ with some non-trivial identification that does {\it not} have a fixed point. For example, identifying $x^3$ with $x^3 +2\pi R$ we get $\mathbb{R}^2 \times \text{S}^1$. This identification does not have a fixed point, and as a result $\mathbb{R}^2 \times \text{S}^1$ is flat. The extensive search  failed to detect any sign of non-trivial classical topology.

There is a good reason why the focus so far has been on classical topologies: If the identification has fixed points, then typically at the fixed points there is a curvature singularity and General Relativity breaks down. This is the sense in which these are non-classical topologies. This seems to suggest that we do not have tools to describe cosmology with non-classical topologies. However, a nice feature of string theory is that in many cases it resolves exactly these kinds of singularities. Roughly speaking, the way this comes about is that in string theory there are excitations   that are confined to the fixed points and, together with the standard excitations that are free to propagate away from the fixed points, provide a consistent description of the physics everywhere, including at the fixed points (see e.g. \cite{polchinski}). From the point of view of string theory such topologies are as legitimate as classical topologies.

This motivates us  to initiate a study of the CMB imprints of such stringy topologies. Our focus here is on the simplest stringy topology -- an orbifold point. The orbifold point is defined by the identification
\be\label{orb}
\b{x}-\b{x}_0\sim -(\b{x}-\b{x}_0).
\ee
This identification has a single fixed point at $\b{x}_0$, which is the location of the orbifold point.

\begin{figure}
\begin{center}
\includegraphics[width=0.4\textwidth]{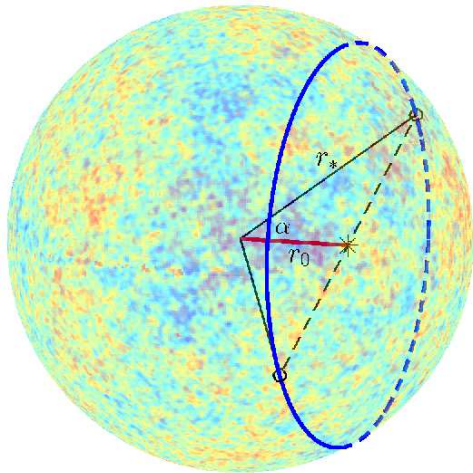}
\caption{An illustration of the points on the last scattering surface (LSS) that are identified with other points on the LSS. As long as the orbifold is within the observable universe, these points form a thin circle. The distance to the orbifold $r_0$ (\emph{thick red line}) and the opening angle of the circle $\alpha$ satisfy $\cos\alpha=r_0/r_*$. The phase separating each matching pair on the circle is $\pi$ (\emph{two small black circles}).}
\label{fig:illustration}
\end{center}
\end{figure}

We do not expect to be able to detect   cosmological imprints associated with  the fixed point itself or the stringy excitations that are confined to it. However, if  $\b{x}_0$ is within the observable universe then it is possible that (\ref{orb}) leaves a detectable imprint on the CMB sky. Part of the orbifold's imprint on the CMB is easy to illustrate. If the orbifold point is within the visible universe,  there will be a special circle in the CMB that will be invariant under rotation by $180^\circ$ (see Fig.~\ref{fig:illustration}).
We refer to such a circle as a Self Matching Circle (SMaC). In this work we study the signal to noise ratio (S/N) associated with a SMaC and more general imprints of an orbifold point on the CMB, and compare it to the data of the Internal Linear Combination (ILC) map of the Wilkinson Microwave Anisotropy Probe (WMAP).

\section{Orbifold Point}

In this section we discuss in some detail the orbifold point. In the first subsection we consider inflation in the presence of an orbifold point. We show that the orbifold does not induce non-Gaussianities. It does, however, modify the power spectrum in a specific way that breaks translation invariance and isotropy.  We use this power spectrum in the second subsection to calculate the CMB anisotropy two point function.

\subsection{Inflation with an Orbifold}

We assume the basic setup of  slow-roll inflation in which the inflaton is  a scalar field $\phi$. Then the orbifold identification (\ref{orb}) implies
\be \label{eq:phiIdent}
\phi(\eta,\b{x})=\phi(\eta,2\b{x}_0 -\b{x}),
\ee
where $\eta$ is conformal time.
Quantum mechanically this means that
\be \label{eq:uComm}
\left[u(\eta,\b{x}),\pi(\eta,\b{x} ')   \right] = i \lb  \delta(\b{x} - \b{x} ') + \delta(\b{x} + \b{x} ' - 2\b{x}_0) \rb ,
\ee
where $u(\eta,\mathbf{x}) = a(\eta)\phi(\eta,\mathbf{x})$ (here $a(\eta)$ is the scale factor) and $\pi(\eta,\b{x} )$ is the canonically conjugate variable.

This yields the following commutation relation for the creation and annihilation operators
\be\label{eq:comm}
\left[a_\b{k} , a_{\b{k} '}^{\dagger}  \right] = (2\pi)^3 \lb
\delta(\b{k} -\b{k} ') + e^{-2i \b{k}\cdot\b{x}_0} \delta(\b{k}+\b{k} ')\rb,
\ee
(and  $[a_\b{k} , a_{\b{k} '}  ] = [a_{\b{k}}^{\dagger} , a_{\b{k} '}^{\dagger}  ]=0$).
The second term is due to the orbifold point and, as expected,  it breaks both homogeneity and isotropy.

Using these commutation relations we find that the theory is still Gaussian. Namely, only the two point functions do not vanish and yield
\be
\label{eq:twoPnt}
\la 0 | \phi(\b{x}) \phi(\b{x} ')  |  0\ra = 
\intk \lb e^{i \b{k}\cdot(\b{x} - \b{x} ')} +
e^{i \b{k}\cdot(\b{x} + \b{x} ' - 2\b{x}_0)} \rb P_{\phi}(k),
\ee
where $P_\phi(k)$ is the standard inflaton power spectrum, $H^2/2k^3,$ evaluated at horizon crossing. Hence for the metric perturbation, $\Phi$,  we get
\be\label{eq:PhiPowSpec}
\la \Phi_\b{k}\Phi_{\b{k}'}^*\ra=(2\pi)^3\left(\delta(\b{k}-\b{k}')+e^{-2i\b{k}\cdot\b{x}_{0}}\delta(\b{k}+\b{k}')\right)P_{\Phi}(k),
\ee
where $P_{\Phi}(k)$ is the standard power spectrum.
Note that violation of translational invariance was considered already in \cite{Carroll:2008br}. The approach taken in \cite{Carroll:2008br} is that this violation is small and the two point function can be expanded around the translational invariant term. In our case  the contributions of the two terms in (\ref{eq:PhiPowSpec}) are of equal magnitudes, and therefore we cannot use such an expansion, but rather use the exact form.

We made two hidden assumptions in deriving (\ref{eq:PhiPowSpec}).
First, we assumed that the string scale is much smaller than the Hubble scale during inflation. Otherwise, the stringy modes that are confined to the orbifold point will affect the inflaton   power spectrum that is fixed at the Hubble scale.
Second, we assumed that the compactifaction scale is smaller than the Hubble distance during inflation. The reason we are forced to make this assumption is that the orbifold point breaks SUSY and as such, leads to instabilities. To avoid this issue  we can consider an orbifold that acts also on the compactified directions (that must exist in string theory) and that on large scales acts like (\ref{orb}). The simplest example is an  orbifold point on $\mathbb{R}^3 \times \text{S}^1$. If the radius of the $\text{S}^1$ is larger than the Hubble scale during inflation then (\ref{eq:PhiPowSpec}) is a good approximation. These assumptions are often made and do not constrain the generality of (\ref{eq:PhiPowSpec}) too much.

\subsection{Angular Correlation Matrix}
\label{sec:AngCorrMat}

Equipped with eq.~(\ref{eq:PhiPowSpec}) the calculation of the harmonic coefficients of the temperature anisotropy is straightforward (though cumbersome) since the $a_{\ell m}$'s are  determined by  $\Phi_{\b{k}}$ in the following way
\be
a_{\ell m}=-\frac{i^\ell}{2\pi^2}\int\d^3 k\, \Phi_{\b{k}}\Delta_{\ell}(k)Y_{\ell m}^{*}(\bhat{k}).
\ee
Here $\Delta_\ell(k)\equiv\Delta_{T\ell}^{(S)}(k,\eta_*)$ is the (scalar) response function and $\eta_*$ is the radius of the last scattering surface (LSS). In using the response function we are utilizing the ``line-of-sight'' approach \cite{Seljak:1996is}, which takes into account the Sachs-Wolfe (SW) effect, the integrated SW effect on large scales and the Boltzmann physics of the coupled photon-baryon fluid on small scales.

We use (\ref{eq:PhiPowSpec}) to calculate the covariance matrix of the harmonic coefficients
\be
C_{\ell m\ell'm'}(\b{x}_{0})\equiv\la a_{\ell m}a_{\ell'm'}^{*}\ra=\delta_{\ell\ell'}\delta_{mm'}C_{\ell}^{(0)}+\Delta C_{\ell m\ell' m'}(\b{x}_{0}), \label{CMat}
\ee
where
\be
C_{\ell}^{(0)}=\frac{2}{\pi}\int k^{2}\d k\,P_\Phi(k)|\Delta_{\ell}(k)|^{2}
\ee
is the standard diagonal \LCDM{} angular power spectrum and
\be\label{eq:departure}
\Delta C_{\ell m\ell'm'}(\b{x}_{0})=i^{\ell+\ell'}\frac{2}{\pi}\int\d^{3}k\,P_\Phi(k)\Delta_{\ell}(k)\Delta_{\ell'}^{*}(k)e^{-2i\b{k}\cdot\b{x}_{0}}Y_{\ell m}^{*}(\bhat{k})Y_{\ell'm'}(\bhat{k})
\ee
is the departure of the correlation function from \LCDM{} due to an orbifold located at $\b{x}_0=r_0\bhat{n}$. In order to evaluate (\ref{eq:departure}) we use the familiar expansion of an exponent in spherical harmonics
\be
e^{i\b{k}\cdot\b{r}}=\sum_{\ell=0}^{\infty}i^{\ell}\sqrt{4\pi(2\ell+1)}j_{\ell}(kr)Y_{\ell 0}(\bhat{k}\cdot\bhat{r}),
\ee
where $j_{\ell}(x)$ is the spherical Bessel function. We fix the coordinate system so that $\bhat{n}$ points in the $\bhat{z}$ direction, and get
\be\label{dc}
\Delta C_{\ell m\ell'm'}(r_{0},\bhat{z})=\frac{2}{\pi}\int k^{2}\d k\,P_\Phi(k)\Delta_{\ell}(k)\Delta_{\ell'}^{*}(k)A_{\ell m\ell'm'}(2kr_{0}),
\ee
where the matrix $A_{\ell m\ell'm'}$ is given by
\be
A_{\ell m\ell'm'}(x)=(-1)^{m}\sum_{\ell''=0}^{\infty}i^{\ell+\ell'-\ell''}\sqrt{4\pi(2\ell''+1)}j_{\ell''}(x)\mathcal{G}_{\ell\ell'\ell''}^{-mm'0}.
\ee
The Gaunt integral $\mathcal{G}$ is defined by
\bea
\mathcal{G}_{\ell_{1}\ell_{2}\ell_{3}}^{m_{1}m_{2}m_{3}} &\equiv&
\int\d^{2}\bhat{k}\,Y_{\ell_{1}m_{1}}(\bhat{k})Y_{\ell_{2}m_{2}}(\bhat{k})Y_{\ell_{3}m_{3}}(\bhat{k}) \nonumber \\
&=&\sqrt{\frac{(2\ell_{1}+1)(2\ell_{2}+1)(2\ell_{3}+1)}{4\pi}}
\begin{pmatrix}
\ell_{1} & \ell_{2} & \ell_{3} \\
0 & 0 & 0
\end{pmatrix}
\begin{pmatrix}
\ell_{1} & \ell_{2} & \ell_{3} \\
m_{1} & m_{2} & m_{3}
\end{pmatrix}
,
\eea
and calculated using Wigner's 3-$j$ symbols. It is non-zero only if $\ell_{1}+\ell_{2}+\ell_{3}$ is even, $m_{1}+m_{2}+m_{3}=0$ and $|\ell_{i}-\ell_{j}| \le \ell_{k} \le \ell_{i}+\ell_{j}$. Therefore, the matrix $A_{\ell m\ell'm'}$ is given by
\be
A_{\ell m\ell'm'}(x)=\delta_{mm'}\sum_{\ell''=|\ell-\ell'|}^{\ell+\ell'}(-1)^{m+(\ell+\ell'-\ell'')/2}\sqrt{4\pi(2\ell''+1)}j_{\ell''}(x)\mathcal{G}_{\ell\ell'\ell''}^{-mm0}.
\ee
The resulting correlation matrix is real, proportional to $\delta_{mm'}$ and symmetric with respect to both $m \leftrightarrow -m$ and $\ell\leftrightarrow\ell'$.

As a consistency check of this equation we take $r_0=0$. Then statistical isotropy is restored and the presence of the orbifold point implies that $C_\ell=0$ for all odd $\ell$'s. Indeed, since $j_{\ell''}(0)=\delta_{\ell''0}$, we get
\bea
A_{\ell m\ell'm'}(0)&=&\delta_{\ell\ell'}\delta_{mm'}(-1)^{\ell+m}\sqrt{4\pi}\mathcal{G}_{\ell\ell0}^{-mm0} \nonumber \\
&=&\delta_{\ell\ell'}\delta_{mm'}(-1)^\ell,
\eea
which immediately gives the diagonal correlation matrix
\be
C_{\ell m\ell'm'}(r_0=0)=\delta_{\ell\ell'}\delta_{mm'}\left[1+(-1)^\ell\right]C_\ell^{(0)},
\ee
as expected.

\section{Detection of an Orbifold in the CMB}

We have seen that the CMB temperature field in a universe with an orbifold point topology differs from that of a trivial topology in its two-point correlation functions. Our goal in this section is to calculate the S/N for the detection of the orbifold 
as a function of its distance $r_0$, in an ideal experiment with no instrumental noise. In the first subsection we  use the full information in the CMB temperature correlation. However, as we explain below, this approach is not feasible. In the second subsection we study a more economical approach that searches for a SMaC in the CMB.

\subsection{Ideal Signal-to-Noise for Detection}

As we have seen, the sole effect of the orbifold point is the deformation of the covariance matrix of the harmonic coefficients without inducing non-Gaussianities. Schematically, we got  $\b{C}_\text{orb}=\b{C}^{(0)}+\Delta\b{C}$, where $\b{C}^{(0)}$ is the standard \LCDM{} covariance matrix. 
Since $\b{C}^{(0)}$ is diagonal (when disregarding instrumental effects), the S/N for detecting the orbifold is \cite{Hamilton:2005kz,Verde:2009tu,Rathaus:2011xi}
\be
\left(\stn\right)^{2}=\text{Tr}\left(\b{C}^{(0)}\b{C}^{-1}-1\right)+\log\left(\det\b{C}/\det\b{C}^{(0)}\right).
\ee
Much of the signal is at high-$\ell$ where this calculation
is extremely computationally intensive. Hence we use the expression expanded to leading order in $\Delta\b{C}$,
\be
\left(\stn\right)^{2}=\sum_{\ell}\left(\frac{\text{S}}{\text{N}}\right)^{2}_{\ell},
\ee
where
\be
\left(\stn\right)^{2}_{\ell}=\sum_{\ell'=2}^{\ell}\left(1-\frac{1}{2}\delta_{\ell\ell'}\right)\sum_{m=-\ell'}^{\ell'}\frac{|\Delta C_{\ell m\ell'm}|^{2}}{C^{(0)}_{\ell}C^{(0)}_{\ell'}}.
\ee
The resulting expression depends only on the distance $r_{0}$ to the orbifold point.

\begin{figure}
\centering
\subfigure[]{\includegraphics[width=0.48\linewidth]{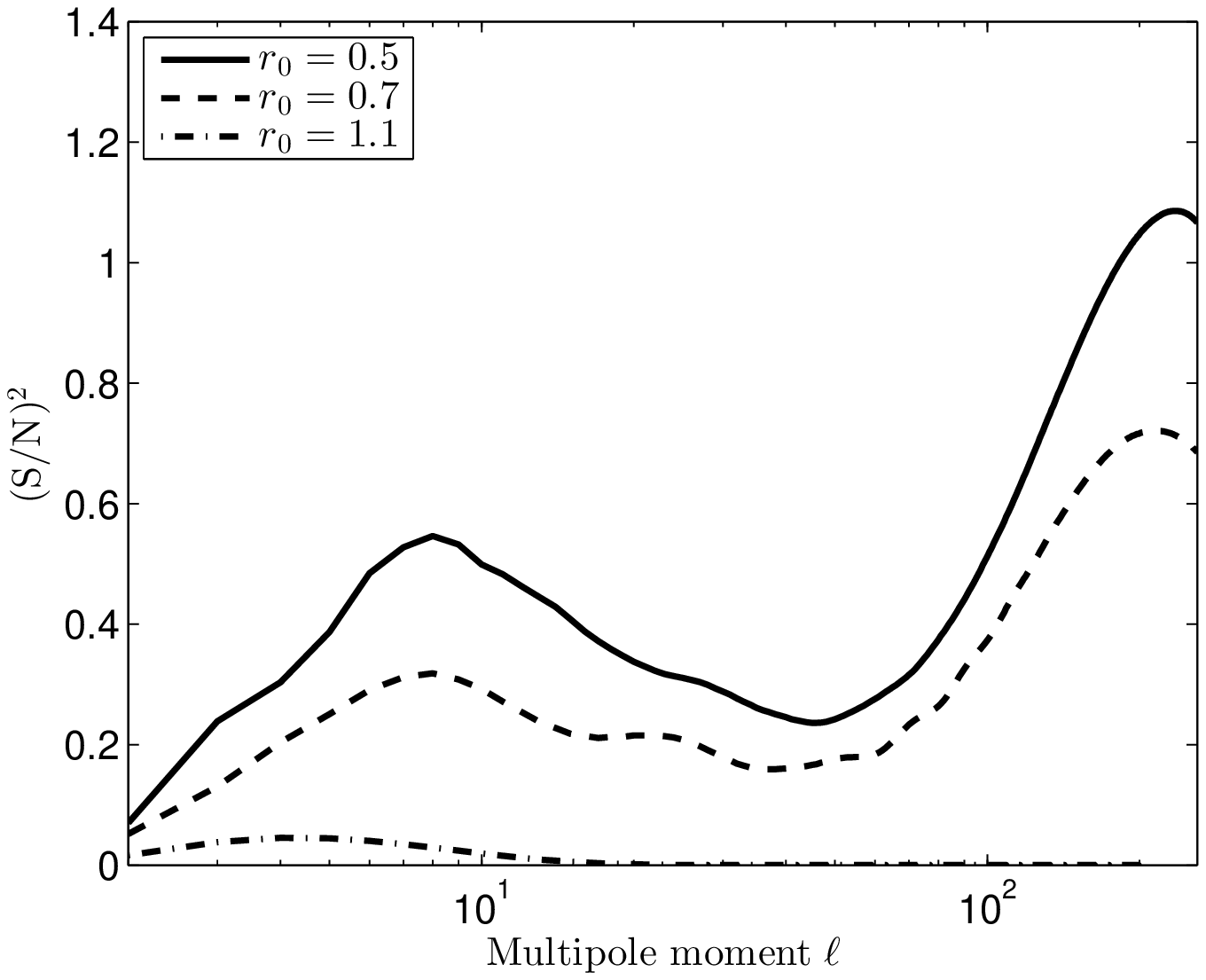}\label{IdealSNPerL}}
\subfigure[]{\includegraphics[width=0.48\linewidth]{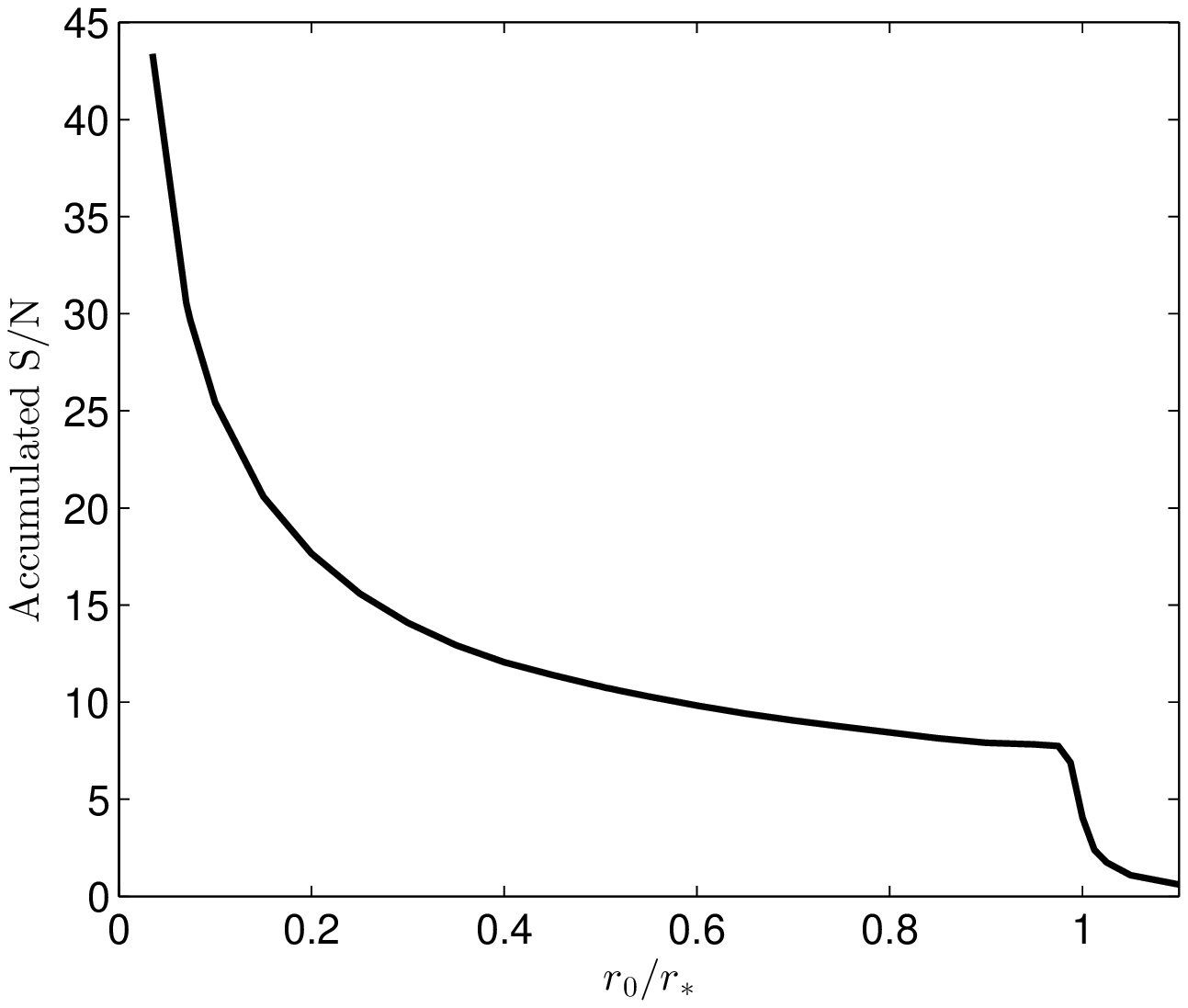}\label{IdealSNTotal}}
\caption{The ideal S/N for orbifold detection, to leading order. (a)~The S/N per multipole $\ell$ for an orbifold located at $r_{0} = 0.5$ (\emph{solid}), 0.7 (\emph{dashed}) and 1.1 (\emph{dot-dashed}), in units of the radius of last scattering. (b)~The accumulated S/N up to $\ell_{\max}=200$ as a function of the distance $r_{0}$.}
\label{IdealSN}
\end{figure}

In Fig.~\ref{IdealSNPerL} we plot the S/N for each multipole $\ell$ and for a few values of $r_{0}$.
We see that as expected the S/N is a decreasing function of $r_0$. This is also apparent from Fig.~\ref{IdealSNTotal}, where the accumulated S/N up to $\ell_{\max}=200$ is shown.\footnote{This scale corresponds to the $1^{\circ}$ resolution of the ILC map.} We can see that the S/N drops fast as the orbifold location approaches the last scattering surface. Once outside, it can no longer be detected. On the other hand, as long as the orbifold is within the observable universe, $\text{S}/\text{N} >5$. There are two contributions to the ideal S/N, one on large scales ($\ell \lesssim 10$) and the other on intermediate scales  ($\ell \sim 200$). The dominant intermediate scales contribution is due to the SMaC and its typical scale is fixed by the thickness of the LSS. In a more realistic setup, the finite beam size is expected to lower the S/N of small scales. We expect, however, that on intermediate scales the effect would be small since the beam size $\lesssim 1^\circ$. In addition the scanning strategy of WMAP is expected to have a direction dependent effect on the S/N, which for a general direction is also expected to be small.

Ideally, we would use the correlation matrix to search for an orbifold signature in the CMB data. In comparing the observed correlations with those of the model, we would utilize as much of the available information as possible. However, we now see that this approach is not feasible, as most of the signal lies in small scales, requiring the analysis of large matrices containing high-$\ell$ data. Since the orbifold model is not isotropic, such an analysis includes the generation of a large correlation matrix for each orbifold distance $r_0$ as well as its rotation to all orientations with respect to the data. We therefore have to search for an orbifold signature in a way that will not use all the available signal, but will be computationally feasible. This is done in the next subsection by focusing on the SMaC only.

\subsection{A Self Matching Circle}

As explained in the introduction, if the orbifold point is within the visible universe, we expect to find a SMaC in the CMB. The opening angle $\alpha$ of the circle satisfies
\be
\cos\alpha = r_0 / r_*,
\ee
where $r_*$ is the radius of the LSS, as illustrated in Fig.~\ref{fig:illustration}.
Every point on this circle is identified with another point on the circle, with a phase of $\pi$ separating the two.
Therefore, the CMB temperatures of the two points match.
If the CMB temperature fluctuations were comprised of the SW signal alone, we would see an exact image of the surface of last scattering and the match would be perfect.  There are, however, other effects that add noise to this match. Experience from the celebrated search for matching circles \cite{Cornish:1997ab,Cornish:2003db,ShapiroKey:2006hm,Bielewicz:2010bh,Vaudrevange:2012da}, relevant for classical topologies, suggests that overall the S/N should be quite high if the circle is large enough. That is, the S/N for detecting the orbifold is expected to decreases with $r_0$. A larger $r_0$ means a smaller circle, and a smaller circle is harder to detect with high statistical significance.

The score for detecting a SMaC is the following. Compare each circle in the data with itself, rotated by a phase of $\pi$. We therefore consider, for each pixel $p$ and for each opening angle $\alpha$, the score
\be\label{eq:estPixelSpace}
\tilde{S}_p(\alpha) = \frac{\la  T_p(\alpha,\phi) T_p(\alpha,\phi+\pi) \ra}{\la  T_p^{2}(\alpha,\phi)  \ra },
\ee
where $T_p(\alpha,\phi)$ is the temperature at phase $\phi$ along the circle centered at $p$ and the angle brackets denote integration over $\phi$. We follow \cite{Cornish:2003db,ShapiroKey:2006hm,Bielewicz:2010bh,Vaudrevange:2012da} and Fourier transform the temperature profile along the circle as $T_p(\alpha,\phi) = \sum_{m}T_{p,m}(\alpha)\exp(im\phi)$. We give  appropriate weights to the different angular scales \cite{ShapiroKey:2006hm} and get
\be \label{eq:circleScore}
S_p(\alpha) = \frac{\sum_{m}(-1)^m m \left|T_{p,m}(\alpha)\right| ^2}{\sum_{n}n \left|T_{p,n}(\alpha) \right| ^2}.
\ee
This score equals unity for a perfect match and for a random circle its expectation value vanishes.

In order to test if the CMB data contain a SMaC, we should examine all pixels $p$ and all angular radii $\alpha$. We use the HEALPix scheme \cite{healpix} to search all directions using a grid with $N_\text{side}=256$, corresponding to a resolution of $\sim 0.25^\circ$. Following \cite{Cornish:2003db}, when collecting data points along each circle, we linearly interpolate values at $2N_\text{side}$ points. For each $\alpha$, we record $S_{\max}(\alpha)$, the highest value of $S_p(\alpha)$ maximized over all pixels $p$. We search all angles in the range $20^\circ \le \alpha \le 90^\circ$ with a resolution of $0.25^\circ$.

\begin{figure}
\centering
\includegraphics[width=0.48\textwidth]{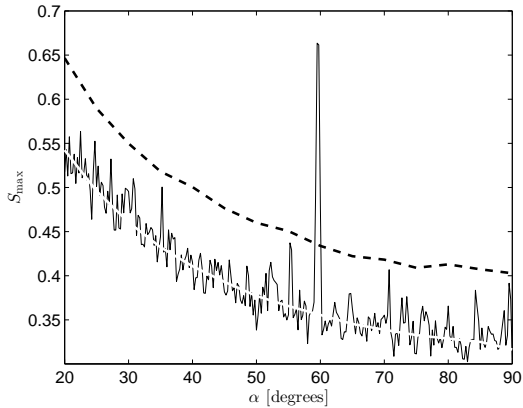}
\caption{The score $S_{\max}(\alpha)$ as a function of opening angle, for a simulated CMB sky with an orbifold located at $r_0=0.5r_*$. A single spike appears at the correct $\alpha$. The dashed line corresponds to the false detection level, discussed below.}
\label{fig:detectionThresholdSingleSim}
\end{figure}

We wish to verify that our score works as it should. To do that, we simulate CMB temperature fluctuation maps for a universe with an orbifold using the covariance matrix, calculated in \S\ref{sec:AngCorrMat}. First, we get the response function $\Delta_{T\ell}^{(S)}(k,\eta_*)$ from CMBFAST \cite{Seljak:1996is} using CMBEASY \cite{Doran:2003sy}, a computation independent of the location of the orbifold. Then, we compute the covariance matrix $\b{C}(r_0)$ of an orbifold located in the $\bhat{z}$ direction. We perform a Cholesky decomposition of the positive-definite matrix to get the triangular matrix $\b{L}$ satisfying $\b{C}=\b{LL}^\dagger$. We then draw a vector $x_{\ell m}$ of uncorrelated complex random Gaussian numbers of zero mean and unit variance, and compute the $a_{\ell m}$ coefficients as $\b{a}=\b{Lx}$. For an orbifold in a different direction, the coefficients can be rotated using a Wigner rotation matrix. Finally, the temperature fluctuation is given by $T(\bhat{n})=\sum_{\ell m}a_{\ell m}Y_{\ell m}(\bhat{n})$. We use $\ell_{\max}=200$ in our simulations. We plot $S_{\max}(\alpha)$ for a simulation with $r_0=0.5r_*$ in Fig.~\ref{fig:detectionThresholdSingleSim}. We find that, as expected, there is a single spike, at the correct opening angle.

\subsection{Orbifold Detection Threshold}

We can use simulated sky maps to determine the maximal distance to an orbifold that could still be detected with our SMaC score, namely the detection threshold. An orbifold located farther away cannot be detected with our score as its signal is obscured by the statistical noise of \LCDM.

\begin{figure}
\centering
\subfigure[]{\label{fig:detectionThresholdNoMask}\includegraphics[width=0.48\textwidth]{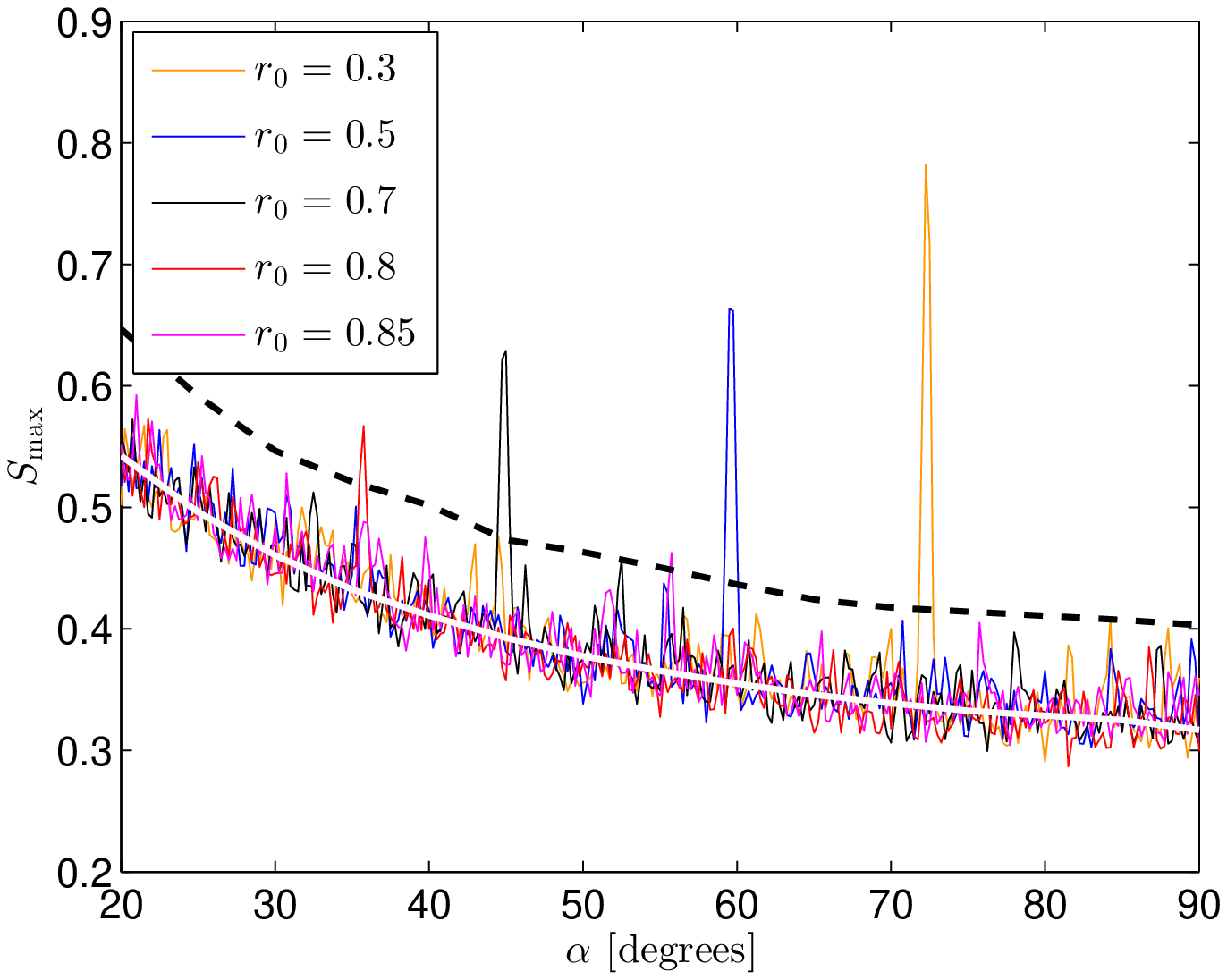} }
\subfigure[]{\label{fig:detectionThresholdMask}\includegraphics[width=0.48\textwidth]{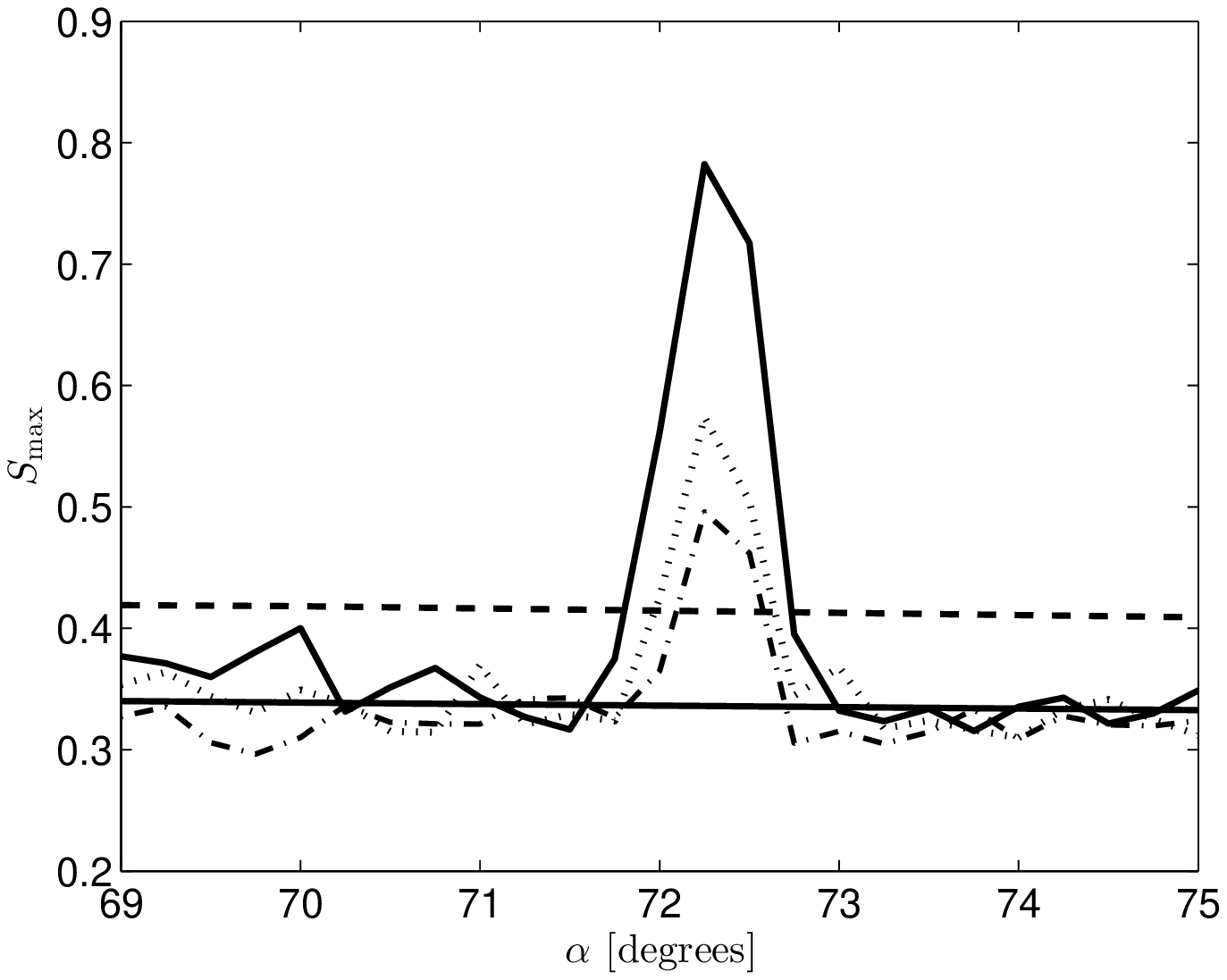} }
\caption{(a) The score $S_{\max}(\alpha)$ for simulated orbifold sky maps at various distances $r_0$. The dashed line is the FDL. As the orbifold is located farther, its peak is less significant, until at $r_0\sim0.85r_*$ it is obscured by the noise. (b) The score for a simulation with an orbifold at $r_0=0.3r_*$ and no contamination (\emph{solid}) and for the same simulation with the data inside the KQ75 galactic mask replaced with a \LCDM{} realization, once when the orbifold is in direction $(l_1, b_1) = (266^\circ, -19^\circ)$ (\emph{dotted}) and once in $(l_2, b_2) = (276^\circ, -1^\circ)$ (\emph{dot-dashed}).}
\label{fig:detectionThreshold}
\end{figure}

To estimate the noise, we simulate \LCDM{} sky maps and calculate $S_{\max}(\alpha)$ for each of them and for each $\alpha$. We then define a false detection level (FDL) corresponding to a level of $3\sigma$ above the median of the score of the random \LCDM{} sky maps.\footnote{Since $S_{\max}(\alpha)$ is not normally distributed we cannot use the mean and standard deviation of the distribution. We therefore use the median and the percentile corresponding to $3\sigma$ instead.}
This is the dashed line shown in Figs.~\ref{fig:detectionThresholdSingleSim},~\ref{fig:detectionThreshold} and \ref{fig:ILCScore}.
It is important to note that the $3\sigma$ FDL we take has the meaning of local significance. By local, we mean the significance of a peak in $S_\text{max}(\alpha)$ with respect to that specific opening angle $\alpha$, and all directions. Of course, once a local peak is found, one needs to take into account the so called ``look elsewhere'' effect to estimate its true significance with respect to all radii and all directions. However, a peak with local significance lower than the FDL should not even be considered as a candidate for further analysis.

We plot $S_{\max}(\alpha)$ for a few values of $r_0$ in Fig.~\ref{fig:detectionThresholdNoMask}.
We can see that an orbifold which is located nearer to us (a SMaC with a larger radius) produces a higher spike in $S_{\max}(\alpha)$, and is therefore easier to detect. In addition, an orbifold located at a distance of $r_0 = 0.85 r_*$  does not produce a peak in $S_\text{max}(\alpha)$ that is significant enough to cross the FDL, setting the detection threshold to $\sim0.85 r_*$.

The above analysis holds for a clean sky. A more realistic scenario, however, must take into account foreground contamination. Even when working with the ILC map, we should still be cautious about pixels that lie inside the galactic mask, as they can affect the matching circle and change the score. For instance, if an orbifold resided in the direction of the galactic north pole at a distance corresponding to, say, $\alpha\sim 80^\circ$, its SMaC would lie entirely inside the mask, risking undetectability even though according to Fig.~\ref{fig:detectionThresholdNoMask} an orbifold at such a location should be easily detected. The mask can also affect smaller circles, with an arbitrary portion of the circle inside the mask.

Since the mask is anisotropic, the FDL is also direction dependent. This in turn means that the detection threshold is direction dependent as well. We demonstrate this by considering two directions that have recently been found to be related to CMB anomalies and examining the dependence of the score on the orientation with respect to the mask. The first is $(l_1,b_1)=(266^\circ,-19^\circ)$, normal to the reflection plane exhibiting large scale odd parity \cite{BenDavid:2011fc,Finelli:2011zs}, and the second is $(l_2,b_2)=(276^\circ,-1^\circ)$, around which \cite{Kovetz:2010kv} found giant concentric rings. In order to generate a simulation with the galactic plane contaminated, we rotate an orbifold simulation so that the orbifold point is in the relevant direction, and replace the data inside the KQ75 galactic mask with those of a random \LCDM{} realization. While this procedure for contaminating the galactic plane obviously does not realistically model any residual galactic foregrounds, it is merely used here to demonstrate the orientation dependence of such effects. After contaminating the same orbifold simulation in orientations suitable for directions $\bhat{n}_1$ and $\bhat{n}_2$, we search for the SMaC with our score. The results, for $r_0=0.3r_*$ as an example, are plotted in Fig.~\ref{fig:detectionThresholdMask}, where it is clear that the orientation affects the level of contamination. Varying $r_0$ and repeating this analysis, we find the detection threshold for these directions. We get $0.35r_*$ and $0.4r_*$ for $\bhat{n}_1$ and $\bhat{n}_2$, respectively, a significant change from the clean-sky isotropic $0.85r_*$.

Unfortunately, calculating an orientation-dependent FDL is not feasible. Therefore, we are left with our isotropic FDL and the naive estimate for the detection threshold. If the data of WMAP show any peaks in $S_{\max}(\alpha)$ that cross the FDL, we should check each candidate and see whether its orientation with respect to the mask is such that it could be heavily contaminated. On the other hand, there is also a chance that a SMaC does exist in the CMB and it is too contaminated for us to detect.

\section{Results}

\begin{figure}
\centering
\subfigure[]{\label{fig:ILCScore}\includegraphics[width=0.48\textwidth]{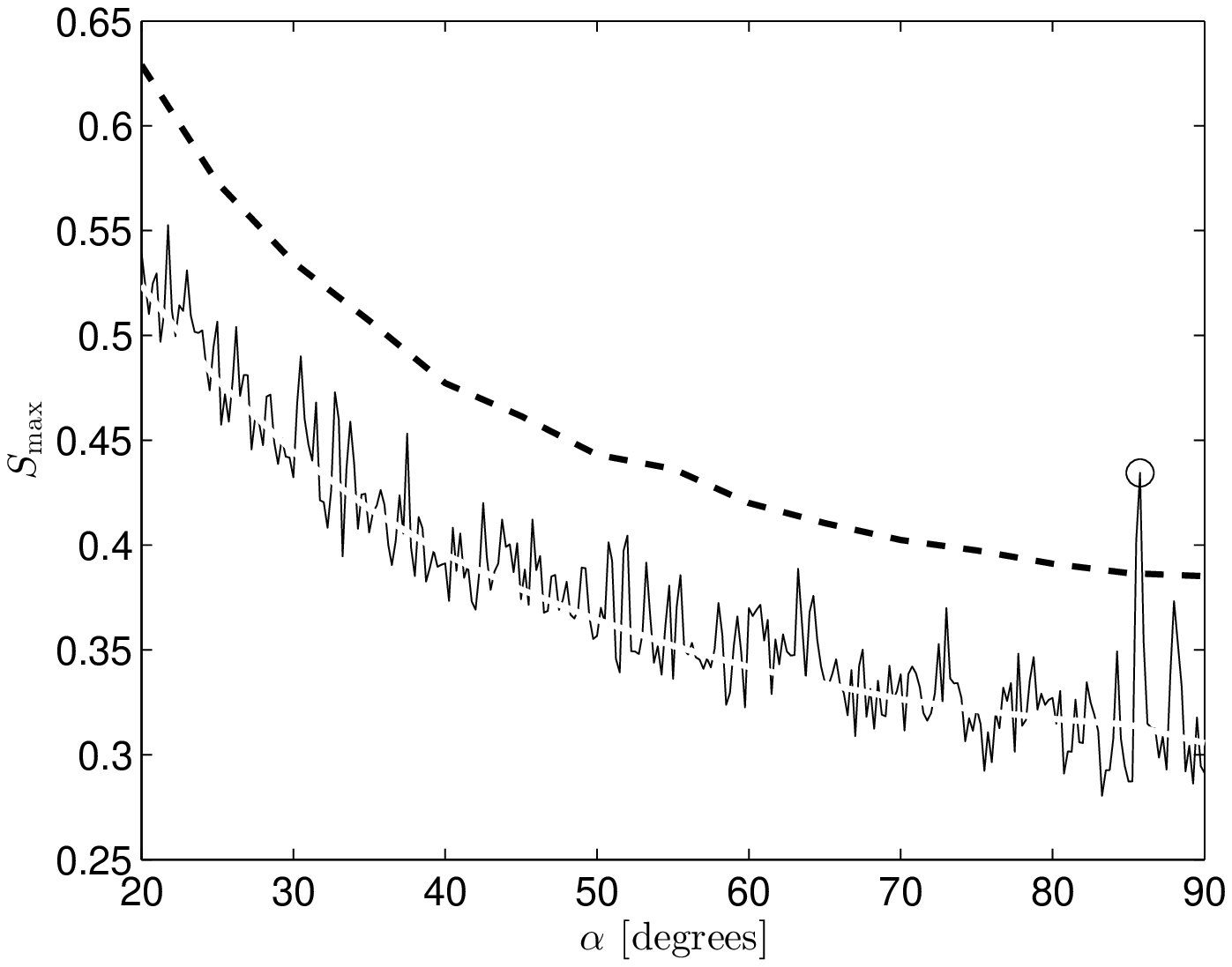}}
\subfigure[]{\label{fig:TTProfileFake}\includegraphics[width=0.48\textwidth]{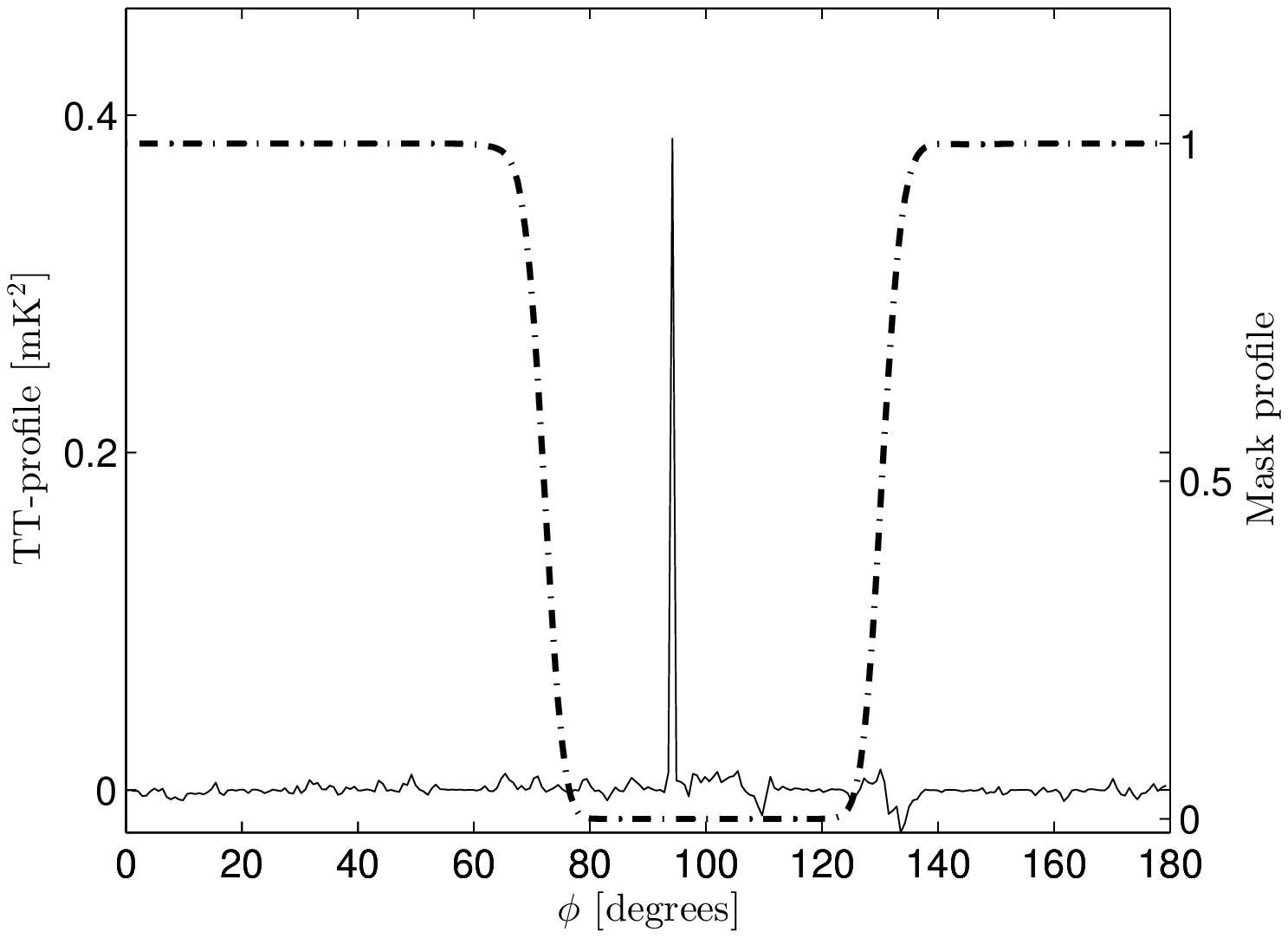}}
\caption{(a)~The score $S_{\max}(\alpha)$ calculated on the 7-year ILC map. The dashed line is the FDL. The peak we analyze further is marked with a circle. (b)~The match profile $T(\phi)T(\phi+\pi)$ as a function of phase $\phi$ for the circle centered at $(l,b)=(275^\circ,-49^\circ)$ with $\alpha=85.75^\circ$ of the ILC map (\emph{solid, left axis}). The (smoothed) KQ75 galactic mask is superimposed (\emph{dot-dashed, right axis}). The entire match of the circle is due to a single pair of pixels, located inside the mask.}
\end{figure}

The score $S_{\max}(\alpha)$ of the WMAP 7-year ILC map is shown in Fig.~\ref{fig:ILCScore}.
It can be seen that one distinct peak crosses the FDL, at $\alpha=85.75^\circ$. It is generated by a circle centered in the direction $(l,b)=(275^\circ,-49^\circ)$. We examine it closer and check if some specific patches in that circle are responsible for its high score. For this sake, we revert back to our score in pixel space (\ref{eq:estPixelSpace}) and plot in Fig.~\ref{fig:TTProfileFake} the match profile $T(\phi)T(\phi+\pi)$ as a function of the phase $\phi$ along the circle. We find that the entire contribution to the score comes from a single pair of matching pixels. Such a profile is extremely inconsistent with the one expected from a SMaC, which implies that this peak is not caused by an orbifold. In addition, we also plot in Fig.~\ref{fig:TTProfileFake} the profile of the KQ75 galactic mask, after smoothing it to $3^\circ$, along the same circle. The anomalous matching pair is well inside the galactic plane, where the mask is zero as indicated in Fig.~\ref{fig:TTProfileFake}. It includes the pixel at $(l,b)=(0,0)$, the center of the galactic bulk and the most heavily contaminated area in the map. It is therefore probable that the peak is a result of some sort of systematic effect or foreground contamination. In any case we can conclude that it does not indicate an orbifold.

Since no other peak crosses the FDL, we conclude that  the ILC data does not show a SMaC, with 99.7\% CL.

\section{Discussion}

Testing the ILC map for a SMaC with our score, we have found no evidence that the universe has an orbifold topology. With the resolution of the ILC map, our score is not sensitive enough to find an orbifold that is located farther than $\sim0.85r_*$.
This, however, does not mean that we can exclude the possibility that there is an orbifold point at $r_0< 0.85r_*$. We have demonstrated how foregrounds can affect the circles, changing their scores, and how important it is to mask the galactic plane. We showed that due to Galactic noise an orbifold point can even be, depending on direction,  as close as  $r_0=0.35 r_*$  without being detected by our SMaC score. Data from Planck, soon to be available, will allow for a better analysis of self-matching circles.

The possibility of finding support for string theory via CMB analysis is exciting. While the topology of the universe remains a mystery, stringy topologies are viable candidates that should be thoroughly explored. We therefore intend to generalize our study here to other stringy topologies.

\acknowledgments
We thank K.~Smith, E.~D.~Kovetz and M.~Kleban for discussions. We acknowledge the use of the Legacy Archive for Microwave Background Data Analysis (LAMBDA) \cite{lambda}. This work is supported in part by the Israel Science Foundation (grant number 1362/08) and by the European Research Council (grant number 203247).

\end{document}